\documentclass[aps,pra,floatfix,twocolumn]{revtex4} 

\usepackage{graphicx}
\usepackage{graphics}
\usepackage{amssymb}
\usepackage{amsmath}
\usepackage{bm}
\usepackage{epsfig}
\usepackage{latexsym}
\usepackage{color}
\usepackage{subfigure}

\newcommand{\tr}{{\rm Tr\thinspace}}
\newcommand{\bra}[1]{\left\langle{#1}\right\vert}
\newcommand{\ket}[1]{\left\vert{#1}\right\rangle}

\newcommand{\abs}[1]{\left\vert #1 \right\vert}
\def\ketc[#1]{\vert #1 \rangle}
\def\brac[#1]{\langle #1 \vert}

\newcommand{\expect}[1]{\left\langle{#1}\right\rangle}

\newcommand{\erf}[1]{Eq.~(\ref{#1})}
\newcommand{\Do}{\Delta\!\omega}
\newcommand{\var}{{\rm var \thinspace}}
\newcommand{\covar}{{\rm cov_\theta \thinspace}}
\newcommand{\dd}[1]{\frac{d}{d#1}}
\newcommand{\dx}[1]{d#1}
\newcommand{\trans}[1]{{#1}^{\ensuremath{\mathsf{T}}}}       
\newcommand{\bth}{{\bm{\theta}}}
\newcommand{\bX}{{\bm{X}}}
\newcommand{\bsig}{{\bm{\sigma}}}

\newcommand{\exper}{\ensuremath{\mathcal{E}}}

\newcommand{\argmax}[1]{\underset{#1}{\operatorname{arg\,max}\,} }
\newcommand{\ifish}{\mathcal{I}}
\newcommand{\id}{{\bm{I}}}

\begin{document}

\title{Optimal quantum multi-parameter estimation and application to \\
	dipole- and exchange-coupled qubits}
\author{Kevin C. Young$^{1}$}
\email{kcyoung@berkeley.edu}
\author{Mohan Sarovar$^{2}$}
\author{Robert Kosut$^{3}$}
\author{K. Birgitta Whaley$^{2}$}
\affiliation{
$^{1}$Berkeley Center for Quantum Information and Computation, Department of Physics, University of California, Berkeley, California 94720, USA \\
$^{2}$Berkeley Center for Quantum Information and Computation, Department of Chemistry, University of California, Berkeley, California 94720, USA \\
$^{3}$SC Solutions, Sunnyvale, California 94085, USA}

\begin{abstract}
We consider the problem of quantum multi-parameter estimation with experimental constraints and formulate the solution in terms of a convex optimization. Specifically, we outline an efficient method to identify the optimal strategy for estimating multiple unknown parameters of a quantum process and apply this method to a realistic example. The example is two electron spin qubits coupled through the dipole and exchange interactions with unknown coupling parameters -- explicitly, the position vector relating the two qubits and the magnitude of the exchange interaction are unknown. This coupling Hamiltonian generates a unitary evolution which, when combined with arbitrary single-qubit operations, produces a universal set of quantum gates. However, the unknown parameters must be known precisely to generate high-fidelity gates. We use the Cram\'er-Rao bound on the variance of a point estimator to construct the optimal series of experiments to estimate these free parameters, and present a complete analysis of the optimal experimental configuration.  Our method of transforming the constrained optimal parameter estimation problem into a convex optimization is powerful and widely applicable to other systems.
\end{abstract}

\maketitle

\section{\label{sec:introduction} Introduction}
The tremendous allure of quantum information processing has fueled recent progress in the experimental and theoretical understanding of physical systems operating in regimes where classical physics fails to hold. The precise control and characterization of physical systems demanded by quantum information processors, e.g. for performing high-fidelity quantum gates, has extended our mastery of optical, gas phase, and condensed phase physical systems. 

Typically, as the precision to which one must characterize a physical system increases, the sophistication of the techniques used to study the system must also increase. Recent \emph{tour-de-force} experiments have fully characterized quantum systems of small dimension by performing exhaustive process tomography (e.g. \cite{OBr.Pry.etal-2004, Wei.Hav.etal-2004}). Such exhaustive tomography requires resources that scale exponentially with the dimension of the system being studied and so is infeasible for systems much larger than those already characterized in this manner. Consequently, many techniques for approximate characterization of large dimensional quantum systems have been formulated in recent years \cite{Moh.Lid-2006, Eme.Sil.etal-2007, Lob.Kor.etal-2008, Kos-2008a}. 

In many situations one is not completely ignorant about the dynamical system being studied. An experimentalist may have partial knowledge of the system through information from system preparation or prior characterization studies. In such cases the system characterization often becomes a problem of parameter estimation, and an important question arises: how does one design an experiment to identify the unknown parameters of the dynamical process most efficiently, or even optimally with respect to some metric? Experiment design for optimal parameter estimation in quantum systems is a natural extension of the equivalent classical design problem; one typically attempts to rapidly reduce the variance in the unknown parameters by performing as few experiments as possible. The goal of experiment design is to identify the input states to probe the dynamical process with and the measurements to perform on the outputs, so that the variance in the unknown parameters can be decreased as quickly as possible (with the number of experiments performed). Analytical and numerical methods for optimal experiment design have been widely explored for one parameter quantum processes (e.g. \cite{Hel-1976, Hol-1982, Bra.Cav-1994, Fuj-2001, Sar.Mil-2006, Boi.Fla.etal-2007}), but very few analytic optimality results exist for the multi-parameter case (for exceptions, see Refs. \cite{Yue.Lax-1973, Fuj.Hir-2003}), in part because of difficulties in optimizing over noncompatibile (noncommuting) quantum observables \cite{Fuj.Nag-1995}. Numerical approaches to optimal experiment design for quantum tomography (when all parameters of the quantum process or state are unknown) and Hamiltonian parameter estimation using convex optimization were first proposed in Ref. \cite{KosutWR-2004}, and applied to experiments in Refs. \cite{Bra.Wal.etal-2007, Bra.Wal.etal-2008}. The method follows from the optimal experiment design approach described in Ch. 7.5 of Ref. \cite{Boy.Van-2004}. Recently, a similar numerical approach to multi-parameter quantum process estimation, using convex optimization, was formulated \cite{Kos-2008} and we shall further refer to it below. Experimentally motivated techniques for multi-parameter estimation have also been proposed \cite{Col.Sch.etal-2005,Dev.Col.etal-2006}, but the optimality and asymptotic performance of these are unknown. 

In this paper we examine this problem of optimal multi-parameter estimation for quantum processes when there are constraints on the possible input probe states and on the possible measurements. The constraints on the input states and spin measurements result from experimental limitations on the types of input states (measurements) that can be realistically prepared (performed). We consider a concrete example motivated by an experimental platform for quantum information processing: donors in semiconductors with electrical control and measurement \cite{Kan-1998, Sch.Per.etal-2003, Sar.You.etal-2007}. We solve the problem of precisely identifying the coupling between two electron spin qubits that interact through a combination of exchange and dipole-dipole interactions by a preparation of input states and measurement of electron spins after a suitable interaction period. Note that precise knowledge of the qubit-qubit interaction is crucial for the execution of two-qubit gates which typically work by transforming this interaction into the desired gate by single qubit manipulation pulses \cite{Zha.Val.etal-2003}. We apply a recently re-formulated numerical approach to optimal
experiment design for multi-parameter quantum estimation
\cite{KosutWR-2004} which also incorporates available experimental
configurations into a convex optimization  \cite{Kos-2008}. This formulation allows us to efficiently identify the optimal characterization experiment and estimate the number of experimental runs necessary to achieve a desired accuracy in the estimated parameters.  

In section \ref{sec:pest_background} we provide background on the quantum parameter estimation problem and recap the formulation of the multi-parameter constrained estimation problem as a convex optimization from Refs. \cite{Kos-2008, KosutWR-2004}. Section \ref{sec:algorithm} presents the experiment design framework in full generality and sketches an algorithm for optimally estimating a set of unknown parameters of a quantum process. Section \ref{sec:physics} introduces the example we explicitly solve in the paper: two coupled electron spin qubits. We summarize the experimental capabilities of this implementation of quantum information processing and give a detailed description of the coupling dynamics. Then in section \ref{sec:results} we apply our experiment design framework to formulate the optimal estimation scheme for identifying the unknown parameters under the given constraints.

\section{\label{sec:pest_background} Parameter Estimation}
Suppose a sequence of data that is independent and identically distributed (\emph{iid}) is drawn from a distribution that is parametrized by one or several unknown quantities.  For instance, the distribution could be Gaussian with unknown mean and variance.  The parameter estimation problem is to estimate the value(s) of the unknown quantities from the sample data.  

A central task of parameter estimation is the construction of an estimator, $T_{\bm{\theta}}(\bm{X})$, which maps the sampled data, $\bm{X}$, to an estimate, $\hat{\bm{\theta}}$, of the parameters.  In what follows, we will assume the use of unbiased estimators,
	\[  \langle\hat\bth\rangle \equiv \expect{ T_{\bm{\theta}}(\bm{X}) } = \bm{\theta}. \]
The generalization to biased estimators is well known, but needlessly complicates our discussion. 

However, some probability distributions are more easily estimable than others.  Take for example a Dirac-delta distribution centered at $x_0$, so that the probability density function is $p_{x_0}(x) = \delta(x-x_0)$.  The parameter to be estimated in this case is $x_0$ and only a single measurement is required.  On the other hand, accurately estimating the mean of a large-variance Gaussian distribution requires many samples.  Estimability of a parameter is thus a property of the probability distribution and is independent of the estimator used.  This idea is encapsulated by the Cram\'er-Rao bound, which places a lower limit on the variance of any single-parameter estimator \cite{Cov.Tho-1991},
	\[ \var T_\theta(\bm{X}) \ge \frac{1}{N F(\theta)}. \]
Here, $N$ is the number of samples and $F(\theta)$ is the Fisher Information, defined as a functional of the probability distribution,
	\[ F(\theta) \equiv \expect{\left( \dd{\theta}   \ln p_\theta(x) \right)^2} \]
where we use the shorthand $p_\theta(x)$ for the conditional distribution $p(x | \theta)$ and $\left<f(x)\right> = \int \!f(x)\,p(x\vert \theta)\,dx$ (or $\sum_i f(x_i) p(x_i\vert\theta)$ if the probability distribution is discrete) is the expectation value of $f(x)$. Note that the Fisher information is a function of the true value (not the estimate) of the parameter. Intuitively it represents the amount of ``information" about the parameter in the conditional probability distribution for the data. 

In the multi-parameter case the generalized Cram\'er-Rao inequality bounds the covariance matrix of the (now vector-valued) estimator \cite{Cov.Tho-1991},
\begin{equation}
	 \covar T_{\bm{ \theta}}(\bm{X}) \ge \frac{\mathcal{I}({\bm{\theta}}\,)^{-1}}{N}
 	\label{eqn:crbound}
\end{equation}
where $\mathcal{I}({\bm{\theta}}\,)$ is the Fisher information matrix:
	\begin{align*}
	 \mathcal{I}(\bm{\theta})
	 	& \equiv 
			\expect{ 
			 \left( \bm{\nabla_\theta}  \ln p_{\bm{\theta}}(x)\right)
			 \trans{\left( \bm{\nabla_\theta}  \ln p_{\bm{\theta}}(x)\right)}
			} 
	\end{align*}
We have used the notation  $\bm{\nabla_\theta} f(\bm{\theta}) = \trans{\left( \dd{\theta_1} f, \dd{\theta_2} f, \ldots, \dd{\theta_n} f \right)}$.  

The Cram\'er-Rao inequality provides a bound on how well we can do when estimating the parameter(s) from the data. While the actual variance in the parameter estimate is dependent on the particular estimator used, there exist estimators that are known to saturate this bound asymptotically (in the limit of large $N$) \cite{Cov.Tho-1991}. An example, that we shall employ below, is the the maximum likelihood estimator (MLE). The MLE is defined as
\begin{equation}
	T^{M\!L}_\bth(\bm{X}) =  \argmax{\bth} p_{\bth}(\bX) 
	\label{eqn:mle}
\end{equation}


\section{\label{sec:algorithm} Optimal Experiment Design for Quantum Parameter Estimation}
Up to this point we have discussed the mathematics of parameter estimation.  The physics of a particular problem becomes important only in calculating the probability distributions (and their derivatives).  Quantum mechanics provides the tools with which these distributions can be obtained.  

We begin by defining an experiment, \exper, as a choice of the initial state, $\rho_0$; evolution time, $t$; and a positive operator valued measure (POVM), ${\bf M} = \{M_i\}$ \cite{Per-1995}. The POVM, also known as a generalized measurement, satisfies $\sum_i M_i = 1$ and $M_i\geq0$, $0\leq i \leq n_{out}$.  Each $M_i$ corresponds to a possible outcome from applying the measurement ${\bf M}$. Through the application of the Born rule, each experiment determines a parametrized family of discrete probability distributions,
	\begin{align}
	p_\bth^\exper(i) 	
		& = \tr\left( M_i \left( U_\bth(t) \rho_0 U_\bth(t)^\dagger \right) \right) \notag \\
		& = \tr\left( M_i \rho_\bth(t) \right)
		\label{eqn:probs}
	\end{align}
Here \mbox{$U_\bth(t) = \mathcal{T}\exp{\left(-i \int_0^t H_\bth(t^\prime) \dx{t^\prime}/\hbar\right)}$} is the unitary time evolution operator and $H_\bth(t)$ is the Hamiltonian whose parameters, $\bth$, we wish to estimate. $p_\bth^\exper(i)$ is the probability, given a fixed experiment $\exper = \{\rho_0, {\bf M}, t\}$ and assuming the parameter takes the value $\bth$, that we get the measurement result $i$. From this probability distribution one can calculate the Fisher information matrix associated to this experiment:
	\begin{equation}
		\mathcal{I}^\exper(\bth) =  \sum_i  \frac{\left( \bm{\nabla_\theta} p_\bth^\exper(i) \right)
							\trans{  \left( \bm{\nabla_\theta} p_\bth^\exper(i) \right)}}
								{p_\bth^\exper(i)}
		\label{eqn:expt_fisher}
	\end{equation}
Inserting this quantity into \erf{eqn:crbound} gives a lower bound on the variance of our estimate.
We will restrict our discussion to closed-system (i.e. Hamiltonian) evolution and, for the sake of clarity, to finite-dimensional Hilbert spaces.  The generalization to non-unitary processes is straightforward; the most difficult step being the calculation of $\bm{\nabla_\theta} p$, which must often be performed numerically.

It is often the case that an experimentalist has access to a number of different initial conditions and measurement bases.  We would like to answer the question: which of these initial conditions and measurements should the experimentalist use in order to best estimate the unknown parameters in the quantum process? In order words, we would like to design our experiment so that we sample the quantum process in a manner that produces the most information about the unknown parameters. Formally. suppose we are given a menu of possible experiments, $\{\exper\}$ and each time we sample our quantum process, an experiment, $\exper = \{\rho_0^\exper, {\bf M}^\exper, t^\exper\}$, is chosen with probability $\lambda_\exper$ (so $\sum_\exper \lambda_\exper=1$).  The result of that measurement is governed by the probability distribution, $p_\bth^\exper(\cdot)$, and so is associated with its own Fisher matrix -- i.e. \erf{eqn:expt_fisher}. The probability of any particular measurement result must now be scaled by the probability that a particular experiment will be performed.  So the Fisher matrix for the combined experimental scheme defined by $\exper$ and $ \lambda_\exper$ is 
	\begin{align*}
	\ifish(\bth) &=\sum_{\exper, i} \frac{\left( \bm{\nabla_\bth}\lambda_\exper p_\bth^\exper(i) \right)
						\trans{\left( \bm{\nabla_\bth}\lambda_\exper p_\bth^\exper(i) \right)}}
							{\lambda_\exper p_\bth^\exper(i)} \\
			&= \sum_\exper \lambda_\exper \ifish^\exper(\bth)
	\end{align*}
It is now natural to ask, given a menu of experiments, what choice of $\bm\lambda$ minimizes $\tr\left(\ifish^{-1}(\bth)\right)=\tr\left(\sum_\exper \lambda_\exper \ifish^\exper(\bth)\right)^{-1}$ and thereby provides the best upper bound on the average of the estimate variance across all parameters through \erf{eqn:crbound}.
This optimization problem, known as \emph{A-optimal experiment design}, can be written as \cite{Boy.Van-2004}:
\begin{align*}
	\rm{minimize} &\hspace{.4cm} \tr \left[ \sum_{\exper} \lambda_\exper \mathcal{I}^\exper(\bth)\right]^{-1} \\
	\rm{subject\, to}&\hspace{.4cm} \lambda_i \geq 0, ~~\sum_i \lambda_i = 1
\end{align*} 
Note that the optimization parameter is the vector of probabilities $\lambda_\exper$. This optimization is difficult because the cost function is not linear or convex in the optimization parameter. However, through the use of the Schur complement (see Appendix \ref{append:schur}), it can be reformulated as the convex optimization problem:
\begin{align}
	\rm{minimize}	&\hspace{.5cm} \tr Q \nonumber\\
	\rm{subject\,to}	&\hspace{.4cm} \left( \begin{array}{cc} Q & I \\ I & F \end{array} \right) \succeq 0,\;
				F =  \sum_{\exper} \lambda_\exper \mathcal{I}^\exper(\bth), \nonumber\\
				&\hspace{.5cm} \lambda_i \geq 0,~~\; \sum_i \lambda_i = 1.
	\label{eqn:optimization}
\end{align}

To make this problem tractable, the menu of experiments can be chosen as a discretization of the continuous space of all possible experiments.  The exact nature of the discretization must be determined for each problem individually, but, in general, a finer grained discretization produces a larger optimization problem.  A coarser graining will result in a smaller optimization problem, but one whose solution will more poorly approximate the true achievable lower bound on the variance given by \erf{eqn:crbound}.  In practice, one will discretize the space of initial states, the space of POVMs, and time.  Given $n_\rho$ initial states, $n_M$ POVM's, and $n_t$ times, we have $\tilde n = n_\rho \times n_M \times n_t$ experiments, and thus an optimization vector, $\lambda_\exper$, of length $\tilde n$.

This procedure for framing the optimal estimation problem as a convex optimization over a discrete space space of experiments is extremely powerful.  Experimental constraints can be used to limit the menu of possible experiments and the optimal distribution can be found quickly, even for large problems. Such a restriction to exclude unfeasible experiments is very difficult to incorporate into a continuous optimization technique. From the convex structure of the optimization, we also gain insight into the expected results.  By the complementary slackness theorem \cite{Boy.Van-2004}, we expect only a small subset of the possible experiments to contribute to the optimal distribution. This expectation is borne out in the example presented in section \ref{sec:results}.

Given this formulation for identifying the optimal experiment, we now detail the entire optimal experiment design procedure:
\begin{description}
\item[ 1. Guess parameters.] We always need an initial estimate of the unknown parameters with which to begin. This assumed value of the parameters, $\bth_p$, can be based on prior knowledge about the quantum process, other studies, or even an educated guess.
\item[ 2. Enumerate possible experiments.] The menu of possible experiments $\exper$ is dictated by experimental constraints. 
\item[ 3. Calculate Fisher matrices.] For each experiment on the menu from step 2, the probability distribution for the outcome data, $p_{\bth_p}^\exper (i)$, and associated Fisher matrices $\mathcal{I}^\exper(\bth_p)$, must be calculated using the assumed value of the parameters.
\item[ 4. Perform optimization.] The optimization specified by \erf{eqn:optimization} must be performed to obtain an optimal probability distribution of experiments, $\lambda_\exper$.
\item[ 5. Perform experiments.] The unknown quantum process should be probed with experiments distributed according to $\lambda_\exper$. That is, if a total of $N$ samples are taken, $\lceil \lambda_\exper/N \rceil$ of them should be using experiment $\exper$.
\item[ 6. Estimate parameter(s).] Use the collected data to estimate the parameters using an estimator of choice. This results in the refined parameter estimate, $\bth_e$. If the maximum likelihood estimator is used, we can readily form the likelihood function since the $N$ samples are independent -- the likelihood function will be a multinomial distribution:
\begin{equation}
	\label{eqn:gen_like}
	p_\bth(\bX)  
		\propto N!
		\prod_\exper \frac{1}{n_\exper !}\prod_{i=1}^{n_{out}^\exper}
			\left(\lambda_\exper p_\bth^\exper(i)\right)^{n_i^\exper} 
\end{equation}
where $n_\exper$ is the number of times experiment $\exper$ was performed ($n_\exper =  \lceil \lambda_\exper/N \rceil$, and $n^\exper_i$ is the number of times result $i$ (corresponding to POVM element $M^\exper_i$) is obtained. The $\bth$ that maximizes this likelihood function is $\bth_e$, the maximum likelihood estimate.
\item[ 7. Repeat if necessary.] This procedure can be repeated, with $\bth_p$ in step 1 replaced by $\bth_e$ from step 6. The decision of whether or not to repeat the procedure can be based on a number of factors: (i) experimental resources, (ii) desired accuracy: if $\bth_e$ is very different from $\bth_p$ then repeating the steps is likely to be helpful. Such an adaptive procedure will converge on the true value of the parameter(s), $\bth_t$, through repetition.
\end{description}

We now illustrate this procedure by treating a specific example of constrained multi-parameter estimation that is very relevant to quantum computing: the identification of coupling parameters in a multi-qubit system.

\section{\label{sec:physics} Dipole- and Exchange-Coupled Qubits}

Donors in silicon have been of increasing interest in the quantum computing community since the seminal paper by Kane in Ref. \cite{Kan-1998}.  Most donor based quantum computing schemes use the spins of electrons bound to donors to encode qubits. Single qubit readout for this implementation is an active area of research, but electrically detected magnetic resonance techniques \cite{Ste.Boe.etal-2006, McC.Hue.etal-2006, Lo.Bok.etal-2007, Sar.You.etal-2007, Mor.McC.etal-2008} are showing potential for delivering high-quality single qubit measurements.  In order to execute high-fidelity quantum gates, accurate knowledge of the coupling Hamiltonian between two donor-bound electron spins is required.  Given exact knowledge of the location of the donors in the substrate, this coupling could in principle be computed theoretically. However, donors in silicon devices are subject to uncertainty in location that is only magnified by subsequent annealing processes. Hence it is highly likely that it will be necessary to characterize the qubit couplings for each device separately and therefore an efficient (and preferably optimal) method of doing this characterization is highly desirable. As we will demonstrate in the next section our constrained parameter estimation scheme is well suited to this task because it is numerically efficient and can handle realistic experimental constraints. Before applying our technique we present some details about the physical system.
 
Two electrons bound to donors implanted in silicon will interact through a combination of the dipole and exchange interactions.  The spin Hamiltonian governing dipole coupling between two qubits is 
\[ H_\textrm{d} = 	  \sum_{i,j}   \gamma_1\gamma_2 
					\expect{  \frac{ 3 \hat{r}_i \hat{r}_j - \delta_{ij}}{4 \pi \abs{\bm r}^3}
							+ \frac{2}{3}\delta_{ij}\delta^3(\bm{r}) }
				         \sigma_i^{(1)} \otimes \sigma_j^{(2)} \]
where $\gamma_i$ is proportionality factor relating the magnetic dipole moment operator to the Pauli matrices:
	\[ \hat{ \bm{\mu} }= \gamma_i \bm{\sigma^{i}}, \]
$\bm{r}= \bm{r}_2-\bm{r}_1$ is the vector connecting the two qubits, and 
$\langle \hat{ \mathcal{O}} \rangle = \bra{\bm{\Psi}}\hat{ \mathcal{O}}\ket{\bm\Psi}$ is the expectation value of $\hat{\mathcal{O}}$ over the two electron spatial wavefunction, $\Psi(\bm{r_1}, \bm{r_2})$. 


The exchange Hamiltonian, a consequence of the Coulomb interaction applied to identical \mbox{spin-$\frac{1}{2}$} particles, is
	\[ H_\textrm{e} = J \bm{\sigma}^{(1)} \cdot \bm{\sigma}^{(2)}. \]
Here, $J$ is the magnitude of the exchange interaction, calculable from the localized, single-qubit wavefunctions, $\phi, \psi$,  by:
	\[ J = e^2 \iint \frac{\phi^*(\bm{r}_1) \psi^*(\bm{r}_2)\phi(\bm{r}_2)\psi(\bm{r}_1)}{\abs{\bm{r}}}
		\textrm{d}\bm{r_1}\textrm{d}\bm{r_2} \]

The qubits are also subjected to a magnetic field, $B = B_0 \bm{\hat z}$, leading to the Hamiltonian,
	\[H_0 = -\gamma_1 B_0 \sigma_z^{(1)}  -\gamma_2 B_0 \sigma_z^{(2)}.  \]
The dipole moments, $\gamma_i$, of the two qubits may be different due to local inhomogeneities in the substrate or the magnetic field.  This results in each qubit, even without any dipole or exchange interaction, having a distinct resonance frequency $\omega_i = \gamma_i B_0/\hbar$, with difference $\Delta \omega = \omega_2 - \omega_1 << \omega_1, \omega_2$.

The static interaction, $H_0$, is presumed to be much larger than either $H_\textrm{d}$ or $H_\textrm{e}$, so it is helpful to work in the interaction picture (also known as the rotating frame) \cite{Sli-1996}.  The effective Hamiltonian in the interaction picture is:
\begin{align}
	H_\textrm{I} 
			=  \sum_{i,j} \Bigg( &\gamma_1 \gamma_2 \expect{
				\frac{\delta_{ij}-3\hat{r}_i\hat{r}_j}{\abs{\bm{r}}^3}
				- \frac{8\pi}{3} \delta^3(\bm{r})\delta_{ij} } \nonumber \\
			&+J \delta_{ij} \Bigg)	\tilde{\sigma}_i^{(1)} \otimes \tilde{\sigma}_j^{(2)} 
			  \label{eqn:intHam}
\end{align}
Where, 
	$ \tilde{\sigma}_i^{(\alpha)} = 
				e^{-i \omega_\alpha \sigma_z^{(\alpha)}t}
					\sigma_i^{(\alpha)}
				e^{i \omega_\alpha \sigma_z^{(\alpha)}t} $,
is the $i^\textrm{th}$ Pauli matrix in the rotating frame of the $\alpha^\textrm{th}$ qubit.  These are, explicitly,
	\begin{align*}
		\tilde{\sigma}_x^{(\alpha)}  
			&= e^{- 2 i \omega_{\alpha} t} \sigma_+^{(\alpha)} 
				\;+\; e^{2 i \omega_{\alpha}t} \sigma_-^{(\alpha)} \\
		\tilde{\sigma}_y^{(\alpha)}  
			&= -i \,e^{- 2 i \omega_{\alpha}t} \sigma_+^{(\alpha)} 
				\;+\; i \,e^{2 i \omega_{\alpha}t} \sigma_-^{(\alpha)} \\
		\tilde{\sigma}_z^{(\alpha)}  
			&= \sigma_z^{(\alpha)}
	\end{align*}
Substituting these into \eqref{eqn:intHam}, we find many terms proportional to $e^{\pm2 i (\omega_1+\omega_2) t}$.  These terms are \emph{very} rapidly oscillating and will average to zero in a short time.  We take the rotating wave approximation and neglect these terms, keeping only those that rotate no faster than $e^{\pm 2 i \Do t}$.  This leaves us with:
	\[ H_\textrm{I} \approx \hbar G \sigma_z^{(1)} \otimes \sigma_z^{(2)}
		+ \left( \hbar F e^{2 i \Do t} \sigma_+^{(1)} \otimes \sigma_-^{(2)}
		+ {\rm{h.c.}} \right) \]
where,
	\begin{align*}
		\hbar F &= 2J - \gamma_1\gamma_2
				\expect{ 
				\frac{\left(1-3\hat{r}_z^2\right)}{\abs{\bm{r}}^3} + \frac{16\pi}{3}\delta^3(\bm{r})}\\
		\hbar G &= \;\,J + \gamma_1\gamma_2
				\expect{ 
				\frac{\left(1-3\hat{r}_z^2\right)}{\abs{\bm{r}}^3} - \frac{8\pi}{3}\delta^3(\bm{r})}
	\end{align*}
In the basis 
	$\{ \ket{\uparrow \uparrow},
		\ket{\uparrow \downarrow},
		\ket{\downarrow \uparrow},
		\ket{\downarrow \downarrow}\}$,
this Hamiltonian can be expressed in matrix form as:
	\[ H_\textrm{I} = 
		\hbar \left( \begin{array}{cccc}
			G & 0 & 0 & 0 \\
			0 & -G & F\, e^{2 i \Do t} & 0 \\
			0 & F\, e^{-2 i \Do t}  & -G & 0 \\
			0 & 0 & 0 & G
		\end{array} \right)
	\]
The unitary evolution operator, \mbox{$U_I(t) = \mathcal{T}\exp{\left(-i \int_0^t H(t^\prime) \dx{t^\prime}/\hbar\right)}$} generated by this Hamiltonian is found to be:
	\begin{widetext}
	\begin{equation}
	   U_I(t) = \left(\begin{array}{cccc}
		e^{-i G t} & 0 & 0 & 0 \\
		0 & e^{-i (\Do - G) t}\left( \cos(\Omega t) - i \Do\sin(\Omega t)/\Omega \right)
			& -i F e^{-i (\Do - G) t}\sin(\Omega t) / \Omega
			& 0 \\
		0 
			& - i F e^{-i (\Do - G) t}\sin(\Omega t) / \Omega
			& e^{-i (\Do - G) t}\left( \cos(\Omega t) + i \Do\sin(\Omega t)/\Omega \right)
			& 0 \\
		0 & 0 & 0 & e^{-i G t} 
		\end{array}\right)
	\label{eqn:unitary} 
	\end{equation}
	\end{widetext}
where we have defined $\Omega = \sqrt{F^2 + \Do^2}$.  From this expression for the time evolution operator, the Fisher information matrices can be computed through \erf{eqn:probs}.


\section{\label{sec:results} Optimal Estimation for Dipole- and Exchange-Coupled Qubits}
In the model described above, the parameters to be estimated are $F$, $G$, and $\Delta\omega$.  To simplify the presentation and for ease of visualization, we assume here that $\Delta\omega$ has been found through standard resonance techniques, and focus our attention on the two remaining parameters. The general technique is of course valid for any number of parameters (within computational constraints).  We choose $\Delta\omega=1$ and take as the true parameter values, $\bth_t = (F_t=1.1, G_t=0.9)$.  We work in units where these parameters are dimensionless.

Realistic experimental constraints for the optimization are that the initial states be easy to prepare and the measurements be experimentally accessible. This is satisfied by assuming that all initial states and POVMs are separable. Both the initial state and the POVM set can be specified by the choice of a Bloch-vector for each of the two qubits.  To discretize the space of initial states and POVMs, the Bloch vectors are chosen from among the 26 unit-norm vectors, $\bm{v}$, of the form 
	\[\bm v = (\alpha \bm{\hat x} + \beta \bm{\hat y} + \gamma \bm{\hat z})
			/\sqrt{\alpha^2+\beta^2+\gamma^2}\] 
where $\alpha, \beta, \gamma \in \{\pm 1,0\}$ are not all zero.  These vectors are illustrated in Fig. (\ref{fig:points})
	\begin{figure}[t]
		\includegraphics[width=6cm]{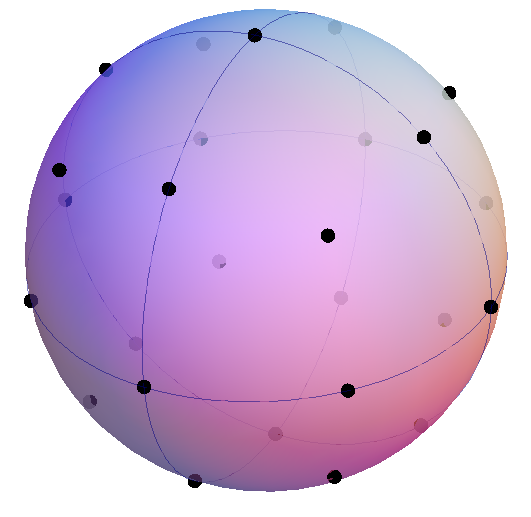}
		\caption{\label{fig:points} (Color online) The single qubit states and measurements used to probe the process represented on the Bloch sphere.}
	\end{figure}

Given a choice of two Bloch vectors $\expect{\bsig^{(1)}_0}$ and $\expect{\bsig^{(2)}_0}$, the density matrix which describes the resulting initial state is 
	\[ \rho_0 = \frac14 \left(\id + \expect{\bsig^{(1)}_0} \cdot {\bsig^{(1)}} \right) \otimes
		\left(\id + \expect{\bsig^{(2)}_0} \cdot {\bsig^{(2)}} \right)\]
Here $\bsig \equiv (\sigma_x, \sigma_y, \sigma_z)$ is a vector formed from the three non-trivial Pauli matrices.
Similarly, given a choice of two Bloch vectors $\expect{\bsig^{(1)}_M}$ and $\expect{\bsig^{(2)}_M}$ the corresponding POVM elements, which we choose in this case projective quantum measurements, are
	\begin{align*}
		M_1 &= \frac14 \left(\id + \expect{\bsig^{(1)}_M} \cdot {\bsig^{(1)}} \right) \otimes
			\left(\id + \expect{\bsig^{(2)}_M} \cdot {\bsig^{(2)}} \right)  \\
		M_2 &= \frac14 \left(\id + \expect{\bsig^{(1)}_M} \cdot {\bsig^{(1)}} \right) \otimes
			\left(\id - \expect{\bsig^{(2)}_M} \cdot {\bsig^{(2)}} \right)  \\
		M_3 &= \frac14 \left(\id - \expect{\bsig^{(1)}_M} \cdot {\bsig^{(1)}} \right) \otimes
			\left(\id + \expect{\bsig^{(2)}_M} \cdot {\bsig^{(2)}} \right)  \\
		M_4 &= \frac14 \left(\id - \expect{\bsig^{(1)}_M} \cdot {\bsig^{(1)}} \right) \otimes
			\left(\id - \expect{\bsig^{(2)}_M} \cdot {\bsig^{(2)}} \right)
	\end{align*}
These are projectors onto anti-podal points (along the axes defined by $\expect{\bsig^{(1)}_0}$ and $\expect{\bsig^{(2)}_0}$) on the Bloch spheres of the two qubits. The set of initial states and POVMs are explicitly enumerated in Appendix \ref{append:opt_expts}. For simplicity, we will fix the duration of each experiment in the menu to $t=1$. Therefore, we have $n_\rho = 26^2, n_M = 13^2, n_t=1$.

The Fisher matrices are calculated using an initial guess $\bth_p= (F_p=1, G_p=1)$ and the optimal experiment is then identified using the convex optimization defined in section \ref{sec:algorithm}. This optimization over $\tilde{n} \equiv n_\rho n_M n_t = 114244$ experiments takes $< 3$ minutes on an average, consumer-grade desktop computer. The result of this optimization is an experimental configuration with only two elements of $\bm{\lambda}^o_\exper > 0$ ($\bm{\lambda}^o_\exper$ is the probability distribution describing the optimal configuration). This means that the optimal process probe need only sample using two experimental configurations out of the $114244$ possible ones. These optimal experimental configurations are shown in Appendix \ref{append:opt_expts}. The non-zero elements of $\bm{\lambda}^o_\exper$ are $0.8$ and $0.2$, which means $4/5$ of the process probes should be performed with one experimental configuration and the remaining $1/5$ with the other. The Fisher information matrix of the optimal configuration is:
\[ \ifish^o(\bm{\theta}_p) = \left(\begin{array}{ccc} 1.8853 & -0.18431 \\-0.18431 & 3.3578\end{array}\right)\]

\begin{figure*}[b]
	\centering
	\subfigure[~Log likelihood function attained using optimal configuration for a single probe time.] 
	{
	    \label{fig:mle:a}
	    \includegraphics[width=5.1cm]{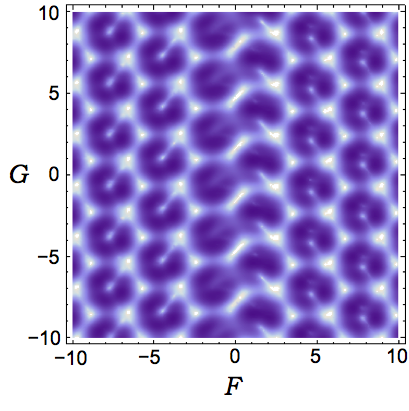}
	}
	\hspace{.5cm}
	\subfigure[~Cross section of the log likelihood function in (a) at the value $G=G_e$.] 
	{
	    \label{fig:mle:b}
	    \includegraphics[width=5.1cm]{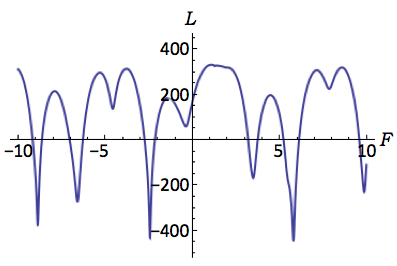}
	}
	\hspace{.5cm}
	\subfigure[~Cross section of the log likelihood function in (a) at the value $F=F_e$.] 
	{
	    \label{fig:mle:c}
	    \includegraphics[width=5.1cm]{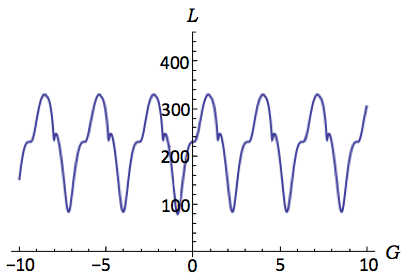}
	}
	\subfigure[~Log likelihood function attained using optimal configuration for multiple probe times.] 
	{
	    \label{fig:mle:d}
	    \includegraphics[width=5.1cm]{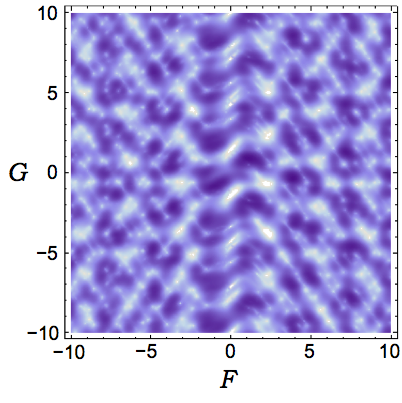}
	}
	\hspace{0.5cm}
	\subfigure[~Cross section of the log likelihood function in (d) at the value $G=G_e$.] 
	{
	    \label{fig:mle:e}
	    \includegraphics[width=5.1cm]{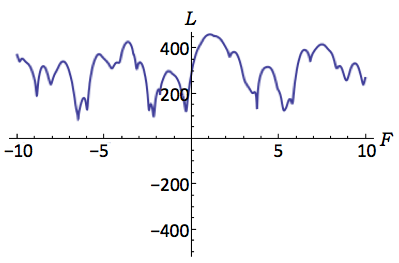}
	}
	\hspace{0.5cm}
	\subfigure[~Cross section of the log likelihood function in (d) at the value $F=F_e$.] 
	{
	    \label{fig:mle:f}
	    \includegraphics[width=5.1cm]{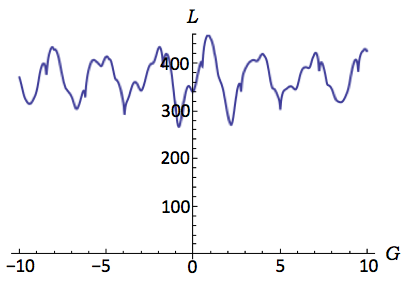}
	}
	\subfigure[~Log likelihood function attained using sub-optimal configuration for a single probe time.] 
	{
	    \label{fig:mle:g}
	    \includegraphics[width=5.1cm]{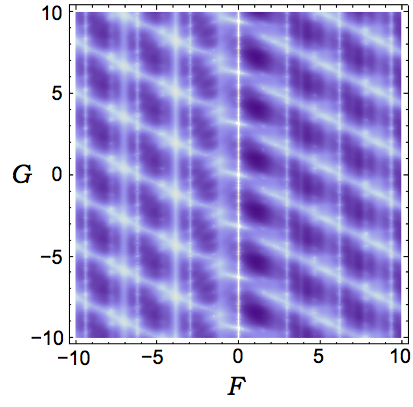}
	}
	\hspace{.5cm}
	\subfigure[~Cross section of the log likelihood function in (g) at the value $G=G_e$.] 
	{
	    \label{fig:mle:h}
	    \includegraphics[width=5.1cm]{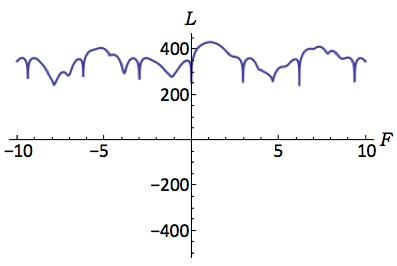}
	}
	\hspace{.5cm}
	\subfigure[~Cross section of the log likelihood function in (g) at the value $F=F_e$.] 
	{
	    \label{fig:mle:i}
	    \includegraphics[width=5.1cm]{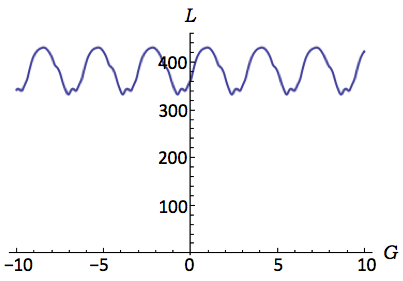}
	}
	\caption{(Color online) The logarithm of the likelihood function, \erf{eqn:gen_like}, for a set of simulated data. Darker areas indicate a larger likelihood function. Sub-figures (a)--(c) show the likelihood function attained using the optimal experiment design. Sub-figures (d)--(f) show the likelihood function attained using the optimal experiment design when the quantum process is probed for different times to break the periodicity of the likelihood in $G$. Figures (g)--(i) show the likelihood function attained using a sub-optimal configuration of experiments to probe the quantum process. In sub-figures (a), (d) and (g) the regions of white along the $F=0$ axis are where the likelihood function is zero and hence its log diverges.}
	\label{fig:mle} 
\end{figure*}

The results of the experiments were simulated using $\bth_t$, the actual value of the parameters and the unitary transformation given by \erf{eqn:unitary}. We sampled the process $N=200$ times with initial states and POVMs dictated by the optimal distribution, $\bm{\lambda}_\exper^{o}$, and the resulting data was used to estimate the parameters using the maximum likelihood estimator, \erf{eqn:mle}. The likelihood function, \erf{eqn:gen_like}, is plotted for a large range of the parameters $F$ and $G$ in Fig. \ref{fig:mle:a}. Finding the maximum over this surface yields $\bth^o_e = (F^o_e=1.10028, G^o_e =0.8845)$. The estimate of the parameters is extremely close to the real values given by $\bth_t$. In addition, we can bound the variance of this estimate using the Cram\'er-Rao bound:
	\[ \textrm{Var}[F^o_e] + \textrm{Var}[G^o_e] \geq \frac{\tr(\ifish^o(\bm{\theta}_p)^{-1})}{200} = 0.0042 \]
As noted earlier, we know that as the number of samples, $N$, increases this bound will be saturated. Thus the estimation error is controlled and well-known. In figures \ref{fig:mle:b} and \ref{fig:mle:c} we show cross sections across the likelihood function at the estimated values $F_e$ and $G_e$. These cross sections show how estimation performance is non-uniform for $F$ and $G$. While the value of $F$ is fairly well resolved, the likelihood function is highly periodic in $G$. This periodicity reflects the periodic manner in which $G$ enters the evolution unitary, \erf{eqn:unitary}. We can break the periodicity of the likelihood function by varying the time for which the quantum process is probed. Figs. \ref{fig:mle:d}--\ref{fig:mle:f} show that likelihood function and its cross sections when the optimal configuration (for $t=1$) is used to probe the process for times $t=1,1.1,1.4$. Again, a total of $N=200$ samples were taken of the process. Periodicity in $G$ is largely absent in \ref{fig:mle:f}, and furthermore, that the likelihood function in \ref{fig:mle:d} has a dominant central peak around the true values of $F$ and $G$. This technique of probing a quantum process for varied times is essential when estimating parameters in unitary processes because of the potential for parameters to appear in a periodic manner in unitary maps. We note that since the probe time, $t$, is actually a parameter of the process it should also be optimized over when identifying the optimal experimental configuration. However, we have not here included this step in the optimization in the interest of keeping the search space of the optimization small enough to explore within $\approx 3$ minutes on our simulation computer. 

To further evaluate the optimal experiment design we compare its performance against a sub-optimal estimation strategy. The sub-optimal strategy we choose is a discrete set of initial preparations and measurements all aligned along the principal Bloch sphere axes ($x,y,z$). The 12 possible experimental configurations for this sub-optimal strategy are listed explicitly in Appendix \ref{append:opt_expts}. This is a reasonable naive strategy, and we again collected $N=200$ samples with experiments distributed uniformly among the 12 possible configurations. The resulting likelihood function is shown in Fig. \ref{fig:mle:g}, and cross sections of it in Figs. \ref{fig:mle:h} and \ref{fig:mle:i}. Taking the maximum over this likelihood surface yields an estimate of the parameters: $\bth^{so}_e = (F^{so}_e=1.085, G^{so}_e=0.969)$. This is clearly a poorer estimate of the true parameters. We can also calculate the Fisher information matrix for this suboptimal strategy:
\[ \ifish^{so}(\bm{\theta}_p) = \left(\begin{array}{ccc} 0.5417 & 0.1662 \\0.1662 & 0.8562\end{array}\right)\]
This Fisher matrix results in the following bound on the combined estimation variance:
	\[ \textrm{Var}[F^{so}_e] + \textrm{Var}[G^{so}_e] \geq \frac{\tr(\ifish^{so}(\bm{\theta}_p)^{-1})}{200} = 0.016 \]
The poorer estimate and the larger bound on the variance for the suboptimal configuration are clear indications of the superiority of the optimal experiment design. Furthermore, the number of experimental configurations required to produce a precise estimate of $\bth$ is vastly smaller for the optimal design.  In Fig. \ref{fig:var_N}, we plot the mean squared error of the maximum likelihood estimate as a function of the number of experiments performed, $N$. While the MLE for both the optimal and sub-optimal configurations approaches the Fisher information bound (provided by the optimal configuration) as $N \rightarrow \infty$, the optimal configuration more rapidly approaches this bound. Furthermore, the mean squared error of the MLE is lower for the optimally configured experiments for all $N$. To achieve the same mean squared error, one must perform roughly twice as many experiments with the suboptimal configuration as are required with the optimal configuration for this particular set of guessed and actual parameters.  

\begin{figure}[h!]
	\centering
	\includegraphics[width=8cm]{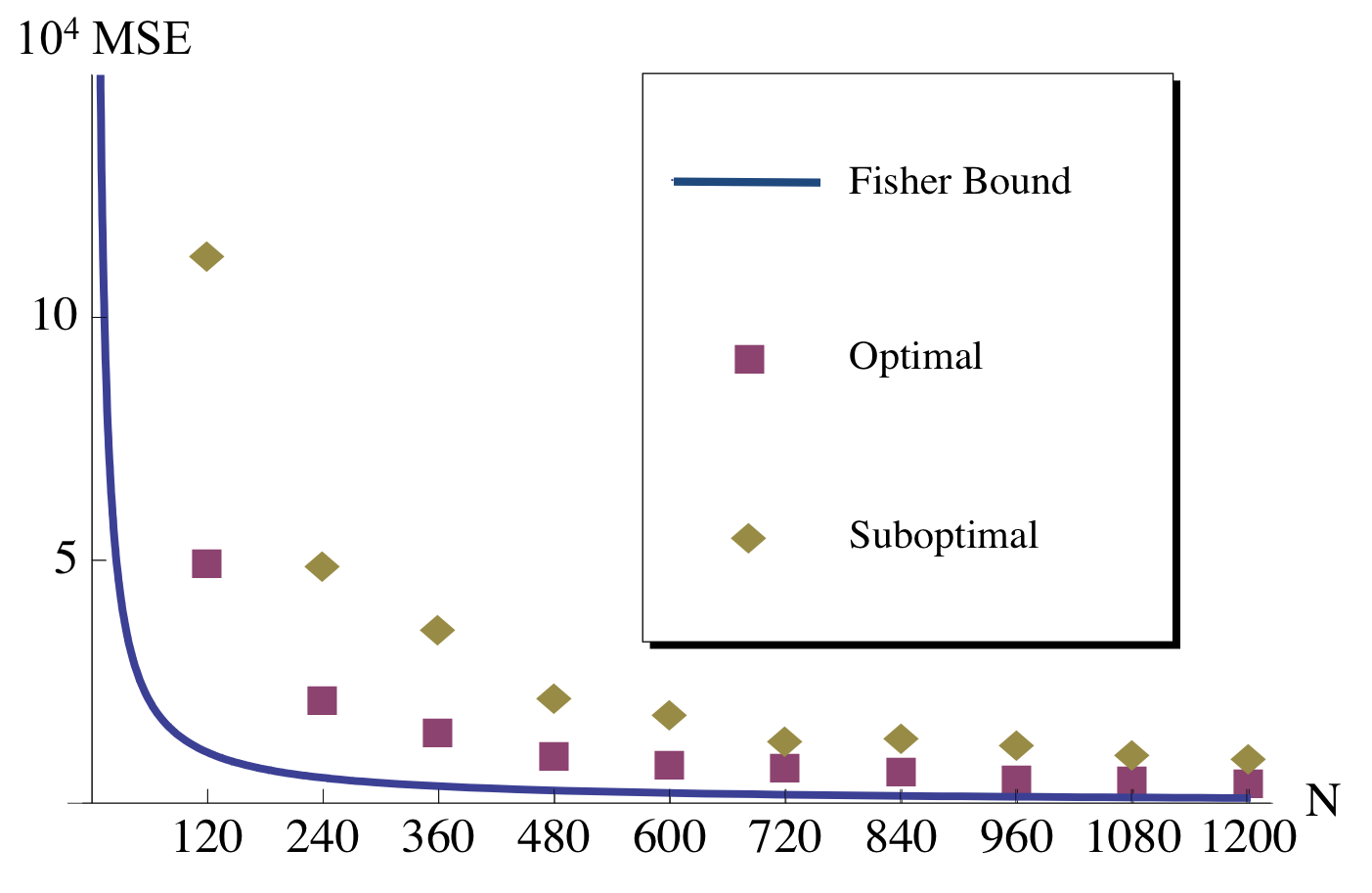}
	\caption{(Color online) Plot of the mean squared error (MSE) of the MLE estimator for the optimal (red squares) and suboptimal (yellow diamonds) configurations.  Also shown (solid blue line) is the Fisher bound for the mean squared error of \emph{any} estimator as given by the optimal experiment.  }
	\label{fig:var_N} 
\end{figure}

To quantify the estimability of the the parameters in this example, we plot the diagonal elements of the inverse of the Fisher information matrix as of function of the parameters, $F$ and $G$ in Fig. \ref{fig:estimability}. Note that the optimal (constrained) probe configuration has been determined for each $(F,G)$ in the plot since this determines the ultimate estimation performance limit.  Fig. \ref{fig:estimability:ff} shows element $[\ifish^{-1}]_{11}(F,G)$ and Fig. \ref{fig:estimability:gg} shows element $[\ifish^{-1}]_{22}(F,G)$. As the values of $F$ and $G$ are changed these elements of the inverse Fisher information matrix remain fairly constant apart from a few notable excursions. This implies that the parameters are nearly equally well estimable across all possible values.  If the optimal probe configuration is utilized the single sample variance bound is $\sim 0.5$ for $F$ and $G$. If the parameters happen to lie on one of the few peaks of $[\ifish^{-1}]_{11}(F,G)$, or $[\ifish^{-1}]_{22}(F,G)$ then they are slightly more difficult to estimate (\emph{i.e.}, a larger number of process probes, $N$, will be necessary to reduce the estimate variance) for any possible experimental configuration. However, note that neither $[\ifish^{-1}]_{11}(F,G)$ nor $[\ifish^{-1}]_{22}(F,G)$ diverge for any value of $(F,G)$, and hence the parameters are always estimable.

\begin{figure}[h!]
	\centering
	\subfigure[~$(1,1)$ element of $\ifish^{-1} (F,G)$] 
	{
	    \label{fig:estimability:ff}
	    \includegraphics[width=7cm]{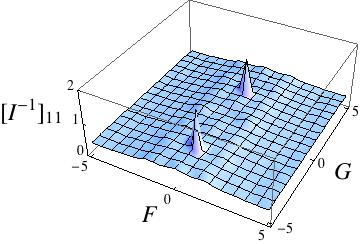}
	}\\
	\subfigure[~$(2,2)$ element of $\ifish^{-1} (F,G)$]
	{
	    \label{fig:estimability:gg}
	    \includegraphics[width=7cm]{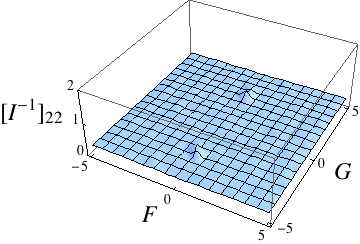}
	}
	\caption{(Color online) The diagonal elements of the optimal inverse Fisher information matrix over a range of values for the unknown parameters, $F$ and $G$.}
	\label{fig:estimability} 
\end{figure}

\subsection{Robustness of Estimation Procedure}

Finally, we turn to the issue of the robustness of the optimal experimental configuration identified by our method. To evaluate robustness, we calculate the inverse Fisher information matrix as a function of the parameters $F$ and $G$ for a fixed experimental configuration (the configuration that is optimal for $(F,G) = (1,1)$). For comparison, we also calculate the inverse Fisher information matrix as a function of the parameters for the fixed sub-optimal process probe configuration used above. The diagonal elements of these matrices are shown in Fig. \ref{fig:robustness}. These figures clearly show that the optimal configuration is much more sensitive to parameter variations than the sub-optimal configuration. In fact, the single sample variance bound for the optimal experiment is quite large at some points.  This is a consequence of the small number of finely tuned experimental configurations utilized by the optimal experiment.  On the other hand, the large number of experimental configurations exploited by the suboptimal experiment allows for a moderate performance for almost all $(F,G)$.  That the optimal experiment is rather sensitive to the accuracy of the initial guess emphasizes the importance of going to the adaptive strategy mentioned in section \ref{sec:algorithm}. That is, as better estimates of the parameters are produced, the process probes should be adapted to be the optimal configurations for the current guess for the parameters. We expect that this lack of robustness of the optimal experiment will be present for the vast majority of parameter estimation problems and is not a special feature of the example considered here. The cost of finely tuning the process probes to optimally estimate the parameters based on an initial guess is that these probes become less adept at identifying values of parameters too far from the initial guess.

Another important point governing the success of the optimization procedure deals with the experimental ability to accurately prepare and measure the qubit states.  In a real experiment, single qubit operations cannot be performed perfectly, and as such will always include a small error.  The state prepared under such a noisy operation will be a mixed state that is proximate to the desired target state.  Such inaccuracies in preparation and measurement can be easily incorporated into our procedure by replacing the probe state (POVM measurement) constellations with the corresponding achievable mixed states (averaged POVMs).  Appendix \ref{append:error} analyzes the specific case of small, random gate error in a single-qubit, single-parameter estimation problem.  The presence of such error is shown to increase the number of experiments required by an amount proportional to a certain measure of the error.

\begin{figure}[h!]
	\centering
	\subfigure[~${[\ifish^{-1}_{11}]}^{\rm Opt}$ Mean:2.03, Min:0.416, Max:27.2] 
	{
	    \label{fig:robust_fish_info:ff}
	    \includegraphics[width=4cm]{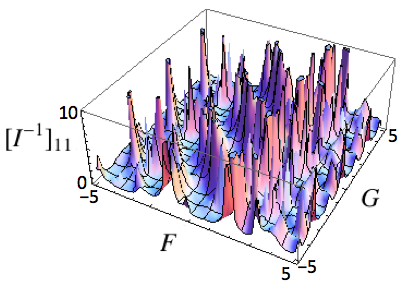}
	}
	\subfigure[~${[\ifish^{-1}_{22}]}^{\rm Opt}$ Mean:0.981, Min:0.300, Max:20.5]
	{
	    \label{fig:robust_fish_info:gg}
	    \includegraphics[width=4cm]{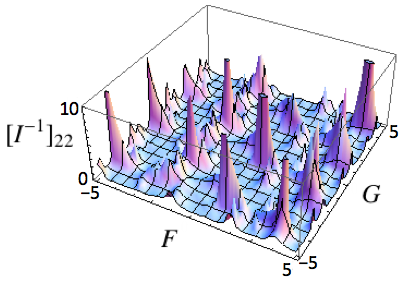}
	}
	\subfigure[~${[\ifish^{-1}_{11}]}^{\rm Sub}$ Mean:1.32, Min:0.234, Max:2.54]
	{
	    \label{fig:robust_subopt_fish_info:ff}
	    \includegraphics[width=4cm]{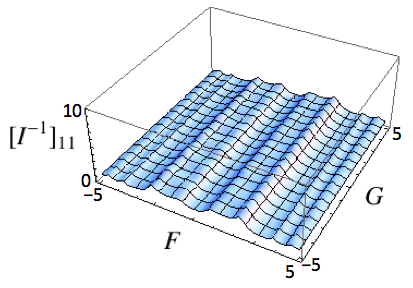}
	}
	\subfigure[~${[\ifish^{-1}_{22}]}^{\rm Sub}$ Mean:1.30, Min:0.213, Max:3.48]
	{
	    \label{fig:robust_subopt_fish_info:gg}
	    \includegraphics[width=4cm]{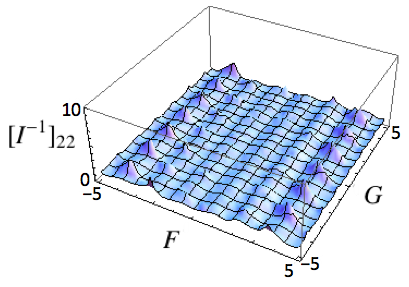}
	}
	\caption{(Color online) The diagonal elements of the inverse Fisher Information matrix over a range of values for the unknown parameters, $F$ and $G$ for fixed process probe configurations. Subfigures (a) and (b) show the diagonal elements of the inverse Fisher matrix when the optimal experimental configuration for the guess $(F,G)=(1,1)$ is used, and subfigures (c) and (d) show these matrix elements when the suboptimal experimental configuration identified in the text is used. As evident from the large deviations in (a) and (b), the small number of experiments used in the optimal configuration reduces the robustness of the procedure to errors in the initial guess. }
	\label{fig:robustness} 
\end{figure}

\section{Conclusion}
\label{sec:conclusion}
The precise estimation of quantum processes is a key ingredient in the engineering of robust quantum information processing devices. For example, to construct two-qubit gates for a quantum computer the interaction between qubits must be precisely known. This estimation task is an increasingly demanding one as the scale of the quantum process being estimated increases. Thus it is essential to have experimental techniques that use minimal resources, but are also accurate. In this work we have demonstrated a method for designing the optimal experiments for multi-parameter quantum process estimation. Particular advantages of the method are that it can tackle multi-parameter estimation, it naturally incorporates realistic experimental constraints, and that the numerical optimization it involves can be implemented efficiently. To demonstrate our approach we have applied it to the estimation of parameters dictating the coupling of two donor electron qubits in silicon. We found the optimal experimental configuration among a very large candidate set ($> 10^5$ experiments) and simulated the parameter estimation using this optimal configuration. The results show that the our method can drastically reduce the number of experiments required to perform parameter estimation for quantum processes. We compared the optimal configuration found by our method with a sub-optimal approach and quantified the performance improvement of the optimal configuration. We also found that while the optimal experiments designed by our procedure -- which are based on an initial guess of the parameters -- perform very well, they are very sensitive to variations in the actual values of the parameters, and hence lack robustness. However, the general algorithm outlined in \ref{sec:algorithm} takes this into account by specifying a recipe for adapting the estimation procedure as data about the values of the parameters is obtained, and hence is capable of compensating for this lack of robustness in the results of the optimization. 

A useful extension of this work is to investigate the feasibility of including a robustness measure directly into the cost function of the optimization.  It remains to be seen if this can be done while maintaining the optimization's convexity.  Robust estimation procedures have been addressed in the context of classical control theory \cite{Lew-2007}  and the extension of these results to quantum models would increase the practicality and appeal of optimal estimation in the quantum setting.


While we illustrated the method here with the example of electron qubits in silicon, the general technique of optimal experiment design for parameter estimation outlined in section \ref{sec:algorithm} is applicable to a wide array of physical systems. An interesting avenue for further research would be to apply this method to identify the Hamiltonians governing small dipole-coupled spin clusters such as those probed in recent experiments with diamond \cite{Neu.Miz.etal-2008, Gur.Chi.etal-2007}.

Finally, although the numerical optimization required to find the optimal experimental configuration is convex, and therefore efficient, in the process of applying our technique to the example detailed above we noticed that current optimization libraries were unable to handle a very large ($\tilde{n} >150,000$) search space. Therefore a possible extension of this work is to use the inherent structure in the parameter estimation problem to form a smaller optimization program, or possibly to iteratively identify the optimal solution.

\appendix


\section{\label{append:schur}Schur Complement}
Consider a nonsingular block matrix,
	\[ M = \left( \begin{array}{cc} A & B \\ C & D \end{array}\right), \]
where $A$ is an invertible submatrix.  The \emph{Schur complement} of M with respect to A is defined as,
	\[ M/A \equiv  D - C A^{-1} B. \]
A principal theorem in the study of the Schur complement \cite{Zha-2005} says that 
	\[ M \ge 0 \iff {\rm{both\;\;}} A \ge 0 {\rm{\;\;and\;\;}} M/A \ge 0. \]
Where $M \ge 0$ means that $M$ is positive semi-definite.  Now consider the minimization problem 
	\begin{align*}
		\rm{minimize}& \hspace{.4cm} \tr F^{-1} \\
		\rm{subject\;to}&\hspace{.4cm} \rm{some\;constraints} 
	\end{align*}
As stated in the main text, this objective function is not convex.  We then propose a matrix $Q \ge F^{-1}$ which is an upper-bound on the matrix $F^{-1}$.  This definition implies that 
	\[ Q-F^{-1} = Q-I F^{-1} I \ge 0.\]
And, because $F$ is positive semi-definite, $F^{-1}$ is as well, implying $Q \ge 0$.  So, by the above theorem,
	\[ \left( \begin{array}{cc} Q & I \\ I & F \end{array}\right) \ge 0. \]
So we construct the following optimization problem:
	\begin{align*}
		\rm{minimize}	& \hspace{.4cm} \tr Q \\
		\rm{subject\;to}	&\hspace{.4cm} \rm{same\;constraints,} \\
					&\hspace{.4cm} \left( \begin{array}{cc} Q & I \\ I & F \end{array}\right) \ge 0.
	\end{align*}
The (convex) matrix positivity constraint enforces the (non-convex) constraint $Q \ge F^{-1}$, leaving us with a convex optimization problem (assuming the remaining constraints are also convex).

\section{Probe constellation and optimal experiments from section \ref{sec:results}}
\label{append:opt_expts}
The 26 single qubit initial states available to probe the quantum process in the example presented in section \ref{sec:results} are given in table \ref{tab:init_states}. Explicitly, for each set of Bloch angles, the initial state is: $\ket{\psi(\phi, \theta)} = \cos(\phi/2)\ket{0}_z + e^{i\theta}\sin(\phi/2)\ket{1}_z$ where $\ket{i}_z$ are $\sigma_z$ eigenstates. The 13 single qubit POVMs assumed available in section \ref{sec:results} are defined by the first 13 angles in table \ref{tab:init_states}, and each POVM has two projector elements given by: $\ket{\psi(\phi, \theta)}\bra{\psi(\phi, \theta)}$ and $\bm{I} - \ket{\psi(\phi, \theta)}\bra{\psi(\phi, \theta)}$.

The sub-optimal estimation strategy used in section \ref{sec:results} used a fixed set of state preparations and measurements to probe the quantum process. There were 12 possible experimental configurations and they are explicitly enumerated in table \ref{tab:suboptimal}. The initial states and POVM elements are defined explicitly in terms of these Bloch sphere angles by the same procedure outline above.

\begin{table}[htdp]
\caption{Bloch sphere angles for the 26 initial states in section \ref{sec:results}. $\phi$ is the polar angle and $\chi \equiv \cos^{-1}(1/\sqrt{3})$.  Antipodal points are equivalent when choosing POVM's, leading to 13 inequivalent, single-qubit measurement bases.}
\begin{center}
\begin{tabular}{|c|c|}
$\phi$ & $\theta$ \\
\hline
0 & 0 \\
$\pi /4$ & $\{0,\pi /2, \pi, 3\pi /2\}$ \\
$\chi$ & $\{\pi/4, 3\pi/4, 5\pi/4, 7\pi/4\}$ \\
$\pi/2$ & $\{0, \pi/4, \pi/2, 3\pi/4, \pi, 5\pi/4, 3\pi/2, 7\pi/4 \}$ \\
$\pi-\chi$ & $\{\pi/4, 3\pi/4, 5\pi/4, 7\pi/4 \}$ \\
$3\pi/4$ & $\{0, \pi/2, \pi, 3\pi/2 \}$ \\
$\pi$ & $0$
\end{tabular}
\end{center}
\label{tab:init_states}
\end{table}

\begin{table}[htdp]
\caption{Bloch sphere angles $(\phi, \theta)$ for the 12 experimental configurations used by the sub-optimal estimation strategy in section \ref{sec:results}. $\phi$ is the polar angle, and Q1 and Q2 refer to qubit 1 and qubit 2.}
\begin{center}
\begin{tabular}{|c|c|c|c|}
Init. state Q1 & Init. state Q2  & POVM Q1 & POVM Q2\\
\hline
$(0,0)$ & $(0,0)$ & $(0,0)$ & $(0,0)$ \\
$(0,0)$ & $(\pi,0)$ & $(0,0)$ & $(0,0)$ \\
$(\pi/2,0)$ & $(-\pi/2,0)$ & $(0,0)$ & $(0,0)$ \\
$(\pi/2,0)$ & $(0,0)$ & $(0,0)$ & $(0,0)$ \\
$(0,0)$ & $(0,0)$ & $(\pi/2,\pi/2)$ & $(\pi/2,\pi/2)$ \\
$(0,0)$ & $(\pi,0)$ & $(\pi/2,\pi/2)$ & $(\pi/2,\pi/2)$ \\
$(\pi/2,0)$ & $(-\pi/2,0)$  & $(\pi/2,\pi/2)$ & $(\pi/2,\pi/2)$ \\
$(\pi/2,0)$ & $(0,0)$ & $(\pi/2,\pi/2)$ & $(\pi/2,\pi/2)$ \\
$(0,0)$ & $(0,0)$ & $(\pi/2,0)$ & $(\pi/2,0)$ \\
$(0,0)$ & $(\pi,0)$ & $(\pi/2,0)$ & $(\pi/2,0)$ \\
$(\pi/2,0)$ & $(-\pi/2,0)$  & $(\pi/2,0)$ & $(\pi/2,0)$ \\
$(\pi/2,0)$ & $(0,0)$ & $(\pi/2,0)$ & $(\pi/2,0)$
\end{tabular}
\end{center}
\label{tab:suboptimal}
\end{table}

The optimal experimental design from the $\tilde{n} = 114244$ possible configurations (defined by all possible combinations of initial state and POVM parameters from Table \ref{tab:init_states}) consists of only two experiments. These are given in Table \ref{tab:optimal}, and graphical representations of the initial states and POVM axes are given in Fig. \ref{fig:optimal_plots}.

Given below are the Fisher information matrices for the experimental configurations chosen by the optimization procedure.  Following these is the inverse of the total Fisher matrix, $ \ifish^{\rm opt}(\bm{\theta}_p) = \sum_\exper \lambda_\exper \ifish_{\lambda_\exper}^{\rm opt}(\bm{\theta})$.
	\begin{align*}
		& \ifish^{{\rm opt}}_{0.2}
			= \left( \begin{array}{cc} 2.03 & -0.034 \\ -0.034 & 2.82 \end{array} \right) \\
		& \left[ \ifish^{\rm opt}_{0.2} \right]^{-1} 
			= \left( \begin{array}{cc} 0.49 & 0.0059 \\ 0.0059 & 0.35 \end{array} \right) \\
		 &\ifish^{\rm opt}_{0.8}
		 	= \left( \begin{array}{cc} 1.85 & -0.22 \\ -0.22 & 3.49 \end{array} \right) \\
		& \left[ \ifish^{\rm opt}_{0.8} \right]^{-1} 
			= \left( \begin{array}{cc} 0.54 & 0.035 \\ 0.035 & 0.29 \end{array} \right) 
	\end{align*}
	\[ \left[ 0.2 \times  \ifish^{{\rm opt}}_{0.2} + 0.8 \times \ifish^{{\rm opt}}_{0.8} \right]^{-1} 
		= \left( \begin{array}{cc} 0.53 & 0.029 \\ 0.029 & 0.30 \end{array} \right) \]

\begin{table}[htdp]
\caption{Bloch sphere angles $(\phi, \theta)$ and relative weights in $\bm{\lambda}^o_\exper$ for the two experimental configurations that are optimal for the estimation problem of section \ref{sec:results}. $\phi$ is the polar angle, and Q1 and Q2 refer respectively to qubit 1 and qubit 2.}
\begin{center}
\begin{tabular}{|c|c|c|c|c|}
$\bm{\lambda}^o_\exper$ & Init. state Q1 & Init. state Q2  & POVM Q1 & POVM Q2\\
\hline
0.2 & $(3\pi/4,3\pi/2)$ & $(\chi,\pi/4)$ & $(\pi/4,0)$ & $(\pi/4,\pi)$ \\
0.8 & $(\pi-\chi,7\pi/4)$ & $(\chi,\pi/4)$ & $(\pi/4,0)$ & $(\chi,5\pi/4)$ \\
\end{tabular}
\end{center}
\label{tab:optimal}
\end{table}

\begin{figure}[t]
	\subfigure[~$\bm{\lambda}^o_\exper = 0.2$ experiment] 
	{
	    \label{fig:optimal_plots:a}
	    \includegraphics[width=5.1cm]{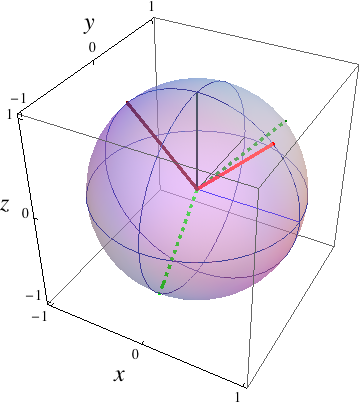}
	}
	\subfigure[~$\bm{\lambda}^o_\exper = 0.8$ experiment]
	{
	    \label{fig:optimal_plots:b}
	    \includegraphics[width=5.1cm]{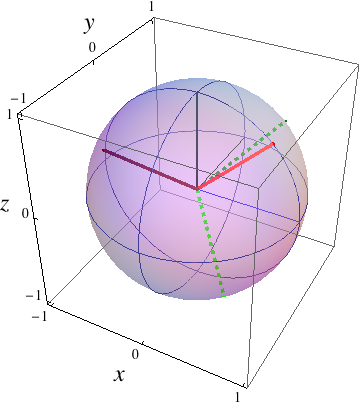}
	}
	\caption{(Color online) Bloch sphere representations of the initial states and POVM axes for the two experiments of the optimal configuration. The green (dotted) lines are Bloch vectors for the initial states of each qubit, and the red (solid) lines define the axes whose antipodal points define the projectors of the optimal POVM for each qubit.}
	\label{fig:optimal_plots} 
\end{figure}

\section{\label{append:error}Robustness to Gate Errors}

When a pure state is acted upon by a noisy gate, the result is a mixed state.
This mixed state can be represented by a Bloch vector which terminates on the
interior of the Bloch sphere.  Though the details depend on the error model for 
the gate, imperfections in preparation and measurement can easily be taken into account by our formalism. One simply optimizes over a discretized set of imperfectly prepared input states and imperfect measurements. Note that even though all inputs states and measurements treated in the example were pure states and projective measurement, respectively, our formalism is not restricted to optimizing over such states and POVMs.
Specifically,  imperfections in state preparation can be taken into account by considering the result of these imperfections on the Bloch vector, $\vec v$, of the state, defined through it's relation to the single-qubit density matrix.  

If we assume that our target state is pure, then $\vec v_f$ is of unit-norm and the density matrix is:
\[ \rho = \frac12 \left( I + \vec v_f \cdot \vec \sigma\right) \]
Random error in preparation of the initial state corresponds to the creation of a mixed state.  If this error is assumed to be such that the final state is instead created with some finite probability density surrounding the target state, the effect on the Bloch vector is that it is contracted by some factor, $\vec v_f^\prime = (1-\epsilon) \vec v_f $.  The details of the error govern the magnitude of $\epsilon$. (Of course, there are errors which do not just shrink the Bloch vector, but also rotate it.  As long as these errors are well characterized, then a similar analysis may be performed.) The density matrix is then,
\[ \rho^\prime = \frac12 \left( I + (1-\epsilon) \vec v_f \cdot \vec \sigma\right) \]
In the case where there exists only a single parameter, $\theta$, the Fisher information takes the form:
\[ \ifish = \sum_i \frac{\left( \frac{\partial}{\partial\theta} p_i(\theta) \right)^2 }{p_i(\theta)} \]
The probability, $p_i(\theta)$, as given by the Born rule for a POVM element, $M_i$, is 
	\begin{align*}
		 p_i(\theta) &= \tr( M_i \rho^\prime) \\
			 	  &= \frac{1}{2} \tr(M_i (I + (1-\epsilon) \vec v_f \cdot \vec \sigma )) \\
				  &= \frac{1}{2} \left( \tr(M_i) + (1-\epsilon)  \tr(\vec v_f \cdot \vec \sigma M_i )\right) 
	\end{align*}
Then the Fisher information becomes, in terms of the Bloch vector,
	\begin{align*}
		 \ifish &= \frac{1}{2} \sum_i 
				\frac{\left( (1-\epsilon) \frac{d}{d\theta} \tr(\vec v_f \cdot \vec \sigma M_i ) \right)^2}
				{  \tr(M_i) + (1-\epsilon)  \tr(\vec v_f \cdot \vec \sigma M_i )}\\
			&\approx (1-\epsilon)^2  \frac{1}{2} \sum_i 
				\frac{\left(\frac{d}{d\theta} \tr(\vec v_f \cdot \vec \sigma M_i ) \right)^2}
				{  \tr(M_i) +  \tr(\vec v_f \cdot \vec \sigma M_i )}\\
			&= (1-\epsilon)^2 \ifish_0 
	\end{align*}
Here, $\ifish_0$, is the Fisher Information achieved without the presence of gate error.  The estimator error, $\var E_\theta$, is bounded by the Cramer-Rao inequality,
\[ \var E_\theta \ge \frac{\ifish^{-1}}{N} = \frac{1}{(1-\epsilon)^2} \frac{\ifish}{N}. \]
So to achieve the same bound on the estimator variance as is found with perfect gates, one must increase the number of measurements from $N$ to $N^\prime \approx N(1+2\epsilon)$.  If there are similar POVM errors as well, then a nearly identical calculation shows that $N^\prime \approx N(1+4\epsilon)$.

\begin{acknowledgments}
This work was supported by the National Security Agency under MOD713100A. The convex optimizations were performed using YALMIP \cite{Lof-2004} and SeDuMi \cite{Stu-2008}. M.S. and K.Y. would like to acknowledge Thomas Schenkel for information about experimental details, and K.Y. is grateful to Yolanda Hagar for useful discussions.
\end{acknowledgments}

\bibliography{./siqc_pest}

\end{document}